\documentclass[aps,prx,print,groupedaddress,twocolumn,,amsmath,amssymb,superscriptaddress]{revtex4-1}
\usepackage{graphicx}
\usepackage{dcolumn}
\usepackage{bm}
\usepackage{subfigure}
\usepackage{color}
\usepackage{txfonts}

\begin{document}

\title{Presence of $s$-wave pairing in Josephson junctions made of twisted ultrathin Bi$_2$Sr$_2$CaCu$_2$O$_{8+x}$ flakes}

\author{Yuying Zhu}\thanks{These authors contributed equally to this work.}
\affiliation{State Key Laboratory of Low Dimensional Quantum Physics and Department of Physics,
Tsinghua University, Beijing 100084, China.}
\affiliation{Beijing Academy of Quantum Information Sciences, Beijing 100193, China.}
\author{Menghan Liao}\thanks{These authors contributed equally to this work.}
\affiliation{State Key Laboratory of Low Dimensional Quantum Physics and Department of Physics,
Tsinghua University, Beijing 100084, China.}
\author{Qinghua Zhang}\thanks{These authors contributed equally to this work.}
\affiliation{Laboratory for Advanced Materials and Electron Microscopy,
Beijing National Laboratory for Condensed Matter Physics, Institute of Physics,
Chinese Academy of Sciences, Beijing 100190, China.}
\author{Hong-Yi Xie}
\affiliation{Beijing Academy of Quantum Information Sciences, Beijing 100193, China.}
\author{Fanqi Meng}
\affiliation{Laboratory for Advanced Materials and Electron Microscopy,
Beijing National Laboratory for Condensed Matter Physics, Institute of Physics,
Chinese Academy of Sciences, Beijing 100190, China.}
\author{Yaowu Liu}
\affiliation{State Key Laboratory of Low Dimensional Quantum Physics and Department of Physics,
Tsinghua University, Beijing 100084, China.}
\author{Zhonghua Bai}
\affiliation{State Key Laboratory of Low Dimensional Quantum Physics and Department of Physics,
Tsinghua University, Beijing 100084, China.}
\author{Shuaihua Ji}
\affiliation{State Key Laboratory of Low Dimensional Quantum Physics and Department of Physics,
Tsinghua University, Beijing 100084, China.}
\affiliation{Frontier Science Center for Quantum Information, Beijing 100084, China.}
\author{Jin Zhang}
\affiliation{State Key Laboratory of Low Dimensional Quantum Physics and Department of Physics,
Tsinghua University, Beijing 100084, China.}
\author{Kaili Jiang}
\affiliation{State Key Laboratory of Low Dimensional Quantum Physics and Department of Physics,
Tsinghua University, Beijing 100084, China.}
\affiliation{Frontier Science Center for Quantum Information, Beijing 100084, China.}
\affiliation{Tsinghua-Foxconn Nanotechnology Research Center, Tsinghua University, Beijing 100084, China}
\author{Ruidan Zhong}
\affiliation{Condensed Matter Physics and Materials Science Department,
Brookhaven National Laboratory, Upton, New York 11973, USA.}
\author{John Schneeloch}
\affiliation{Condensed Matter Physics and Materials Science Department,
Brookhaven National Laboratory, Upton, New York 11973, USA.}
\affiliation{Department of Physics and Astronomy, Stony Brook University, Stony Brook, New York 11794, USA.}
\author{Genda Gu}
\affiliation{Condensed Matter Physics and Materials Science Department,
Brookhaven National Laboratory, Upton, New York 11973, USA.}
\author{Lin Gu}
\affiliation{Laboratory for Advanced Materials and Electron Microscopy,
Beijing National Laboratory for Condensed Matter Physics,
Institute of Physics, Chinese Academy of Sciences, Beijing 100190, China.}
\author{Xucun Ma}
\affiliation{State Key Laboratory of Low Dimensional Quantum Physics and Department of Physics,
Tsinghua University, Beijing 100084, China.}
\affiliation{Beijing Academy of Quantum Information Sciences, Beijing 100193, China.}
\affiliation{Frontier Science Center for Quantum Information, Beijing 100084, China.}
\author{Ding Zhang}
\email{dingzhang@mail.tsinghua.edu.cn}\affiliation{State Key Laboratory of Low Dimensional
Quantum Physics and Department of Physics, Tsinghua University, Beijing 100084, China.}
\affiliation{Beijing Academy of Quantum Information Sciences, Beijing 100193, China.}
\affiliation{Frontier Science Center for Quantum Information, Beijing 100084, China.}
\affiliation{RIKEN Center for Emergent Matter Science (CEMS), Wako, Saitama 351-0198, Japan.}
\author{Qi-Kun Xue}
\email{qkxue@mail.tsinghua.edu.cn}
\affiliation{State Key Laboratory of Low Dimensional
Quantum Physics and Department of Physics, Tsinghua University, Beijing 100084, China.}
\affiliation{Beijing Academy of Quantum Information Sciences, Beijing 100193, China.}
\affiliation{Frontier Science Center for Quantum Information, Beijing 100084, China.}
\affiliation{Southern University of Science and Technology, Shenzhen 518055, China.}
\date{\today}

\begin{abstract}
Since the discovery of high temperature superconductivity in cuprates, Josephson junction based phase-sensitive experiments are believed and used to provide the most convincing evidence for determining the pairing symmetry. Regardless of different junction materials and geometries used, quantum tunneling involved in these experiments is essentially a nanoscale process, and thus, actual experimental results are extremely sensitive to atomic details of the junction structures. The situation has led to controversial results as to the nature of the pairing symmetry of cuprates: while in-plane junction experiments generally support $d$-wave pairing symmetry, those based on out-of-plane ($c$-axis) Josephson junctions between two rotated cuprate blocks favor $s$-wave pairing. In this work, we revisit the $c$-axis experiment by fabricating Josephson junctions with atomic-level control in their interface structure. We fabricate over 90 junctions of ultrathin Bi$_2$Sr$_2$CaCu$_2$O$_{8+x}$ (BSCCO) flakes by state-of-the-art exfoliation technique and obtain atomically flat junction interfaces in the whole junction regions as characterized by high resolution transmission electron microscopy (TEM). Notably, the resultant uniform junctions at various twist angles all exhibit a single tunneling branch behavior, suggesting that only the first half a unit cell on both sides of the twisted flakes is involved in Josephson tunneling process. With such well-defined geometry/structure and the characteristic single tunneling branch, we repeatedly observe Josephson tunneling at a nominal twist angle of 45 degrees, which is against the expectation from a purely $d$-wave pairing scenario. Our results strongly favor the scenario of a persistent $s$-wave order parameter in the junction.
\end{abstract}

\maketitle

\section{INTRODUCTION}
The mechanism of high-$T_c$ superconductivity in cuprates has been a lastingly vibrant topic ever since the advent of cuprate superconductors more than three decades ago~\cite{Keimer_2015,Comin_2016,Hamidian_2016,Chen_2019,Kleiner_2019,Oles_2019,Massee_2020}. By now, the $d$-wave pairing symmetry of cuprates seems to be widely recognized, which is established largely based on the in-plane Josephson junction experiments~\cite{Hilgenkamp_2002}. For instance, cuprate grain boundary junctions~\cite{Lombardi_2002} as well as hetero-junctions between the cuprate and a conventional $s$-wave superconductor~\cite{Smilde_2005} exhibited a modulated Josephson current as a function of the crystal orientation, while superconducting quantum interference experiment of Pb-YBa$_2$Cu$_3$O$_7$ junctions~\cite{Wollman_1993} revealed a phase shift of $\pi$. Local susceptibility measurements on cuprate tricrystals~\cite{Tsuei_1994,Tsuei_2000} and corner/ring junctions of Nb/Au/YBa$_2$Cu$_3$O$_7$~\cite{Hilgenkamp_2003,Kirtley_2006} show spontaneous supercurrent induced magnetic flux. These results have served as strong evidence for the scenario of $d$-wave order parameter that changes sign.

However, there exist also experiments that support the $s$-wave pairing or a mixture of $s$ and $d$-wave pairing~\cite{Mueller_2017}, particularly those with the $c$-axis Josephson junctions~\cite{Li_1999,Takano_2002,Latyshev_2004,Klemm_2005}. Because the Josephson tunneling is extremely sensitive to the detailed structure of the junctions, an obvious advantage using the $c$-axis junctions, compared with the in-plane ones, lies in the relatively well-controlled interface structure that is established by just twisting the freshly cleaved cuprate crystals, in a way analogous to making twisted bilayer graphene~\cite{Cao_2018a,Cao_2018b}. Taking BSCCO as an example, the superconducting CuO$_2$ planes across the BiO/SrO barrier layers can form Josephson coupling~\cite{Kleiner_1992,Kleiner_1994,Krasnov_2000,Suzuki_2000,Suzuki_2012} when stacking two twisted BSCCO crystals together~\cite{Li_1999,Takano_2002,Latyshev_2004}. This process usually leads to atomically flat junction interface, without the faceting problem that is usually unavoidable in fabricating the in-plane junctions~\cite{Zhang_1996,Jin_2002,Hilgenkamp_2002,Arnold_2006}. Theoretically, a purely $d$-wave order parameter demands that the Josephson coupling between two CuO$_2$ planes across the BiO/SrO barrier vanishes if one CuO$_2$ plane is rotated by 45$^\circ$ against the other~\cite{Klemm_1998,Arnold_2000,Bille_2001,Klemm_2003,Klemm_2005,Yokoyama_2007}. In contrast, the Josephson coupling would show no such effect if the superconducting CuO$_2$ planes have $s$-wave pairing symmetry. The angular dependence of the Josephson current in the $c$-axis twist junction would therefore be a litmus test of the $d$-wave pairing symmetry. Experimentally, a pioneer work by using twisted BSCCO bulk crystals was conducted to determine the pairing symmetry based on the theory, but found no angular dependence~\cite{Li_1999}, implying an unusual $s$-wave pairing symmetry~\cite{Arnold_2000,Bille_2001}. A problem in the experiment was from the bulky samples ($\sim10^4$-$10^5\mu$m$^2$ in area), in which the current-induced over-heating only allows the angular dependence comparison conducted at high temperatures, as high as 90$\%$ of the transition temperature, i.e. $T\sim0.9T_c$. Two follow-up experiments with smaller samples (whiskers) were carried out later on and they showed contrasting results~\cite{Takano_2002,Latyshev_2004}. Nevertheless, the results can basically be explained in the frame work of $s$-wave pairing~\cite{Klemm_2003,Klemm_2005}. A common feature in both experiments is the presence of multiple tunneling branches indicative of involvement of the intrinsic multiple junctions in the bulk, which pose difficulty in isolating the contribution of the artificial junction from that of the bulk ones. Furthermore, the inhomogeneities with bulky samples, for example, the presence of Bi-2223 phase, may lead to overestimated Josephson currents at certain angles~\cite{Klemm_2003}. In general, despite its potentially profound implications, the $c$-axis tunneling experiments were not widely accepted by the community due to the above-mentioned uncertainties that are associated with the detailed geometry and structure of the junctions~\cite{Zhu_1998}. It is clear that, without knowing the atomic structure of the whole junctions in actual tunneling, the data interpretation can be challenging and may be less convincing.

Recently emerged technique of van der Waals (vdW) stacking of two-dimensional materials~\cite{Wang_2013,Gomez_2014,Masubuchi_2018}, which has also been employed to successfully address superconductor-insulator transition~\cite{Liao_2018}, dimensionality~\cite{Jiang_2014,Zhao_2019,Yu_2019} and proximity effects~\cite{Zareapour_2012} in BSCCO, provides an unprecedented opportunity to solve this problem. In this work, we adopt this technique and fabricate over 90 twisted bicrystal BSCCO Josephson junctions. Using high resolution TEM, we demonstrate that the junctions fabricated by this method all exhibit atomically sharp interfaces, and more importantly, that this type of interface extends uniformly over the entire tunneling region of the junctions. We further demonstrate that the critical current $I_c$ is solely from the tunneling between the bottom half unit cell~(UC) of the top flake and the top half UC of the bottom flake that constitute the junction, which avoids possible complications from the intrinsic junctions inside the flakes. By carrying out tunneling experiments in these atomically controlled junctions and comparing the data to the theoretical models~\cite{Klemm_1998,Arnold_2000,Bille_2001,Klemm_2003,Klemm_2005,Yokoyama_2007}, we demonstrate that it is necessary to involve a $s$-wave pairing component in the order parameter to explain the persistent Josephson coupling at the twist angle of 45 degrees.

\section{Experiment}
\begin{figure}
\includegraphics[width=86mm]{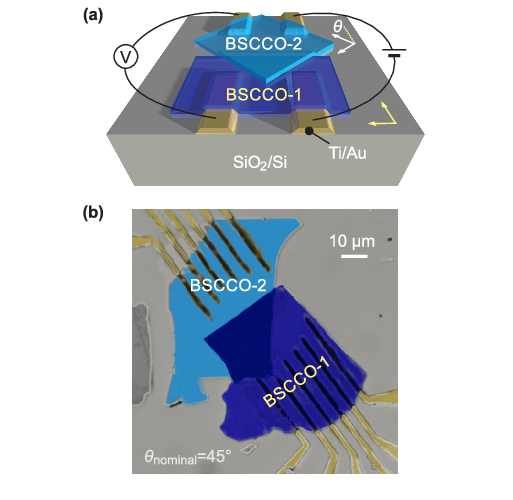}
\caption{High-$T_c$ Josephson junction.
(a) Top: schematic drawing of the Josephson junction made of two overlapping BSCCO flakes on a SiO$_2$/Si substrate. (b) False color image of a 45$^\circ$ twist junction with pre-patterned Ti/Au electrodes taken by an optical microscope. Each flake has a thickness of about 10 unit cells ($\sim$30 nm).}
\end{figure}

\begin{figure*}
\includegraphics[width=178mm]{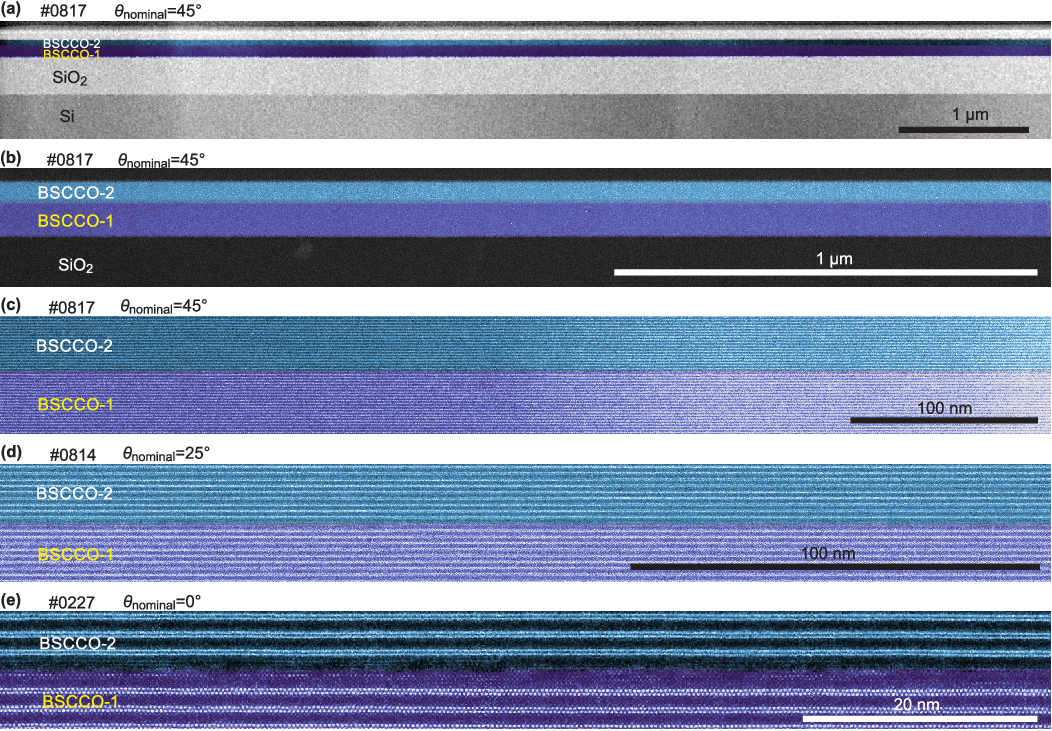}
\caption{
(a) Cross-sectional TEM image of one sample over the entire region cut out by focused ion beam (FIB). (b) TEM image spanning about 2.5~$\mu$m wide on the same sample shown in (a). (c) to (e) TEM images of the junctions with different twist angles illustrating the great uniformity of the samples. The horizontal range is approximately 560 nm for (c), 280 nm for (d) and 100 nm for (e).}
\end{figure*}

Bulk single crystals of nearly optimally doped BSCCO were grown by the traveling floating zone method. They were mechanically exfoliated to obtain ultrathin flakes in a glovebox in Ar atmosphere (H$_2$O$<0.1$~ppm, O$_2$$<0.1$~ppm). Two ultrathin BSCCO flakes (6 to 20~UC thick) were sequentially placed on top of the SiO$_2$/Si substrate with pre-patterned electrodes (Ti/Au: 5/30~nm)~\cite{Liao_2018}. They were rotated against each other as shown in Fig.~1. The designated twist angle was realized by handling scotch tapes and Gel-Films from Gel-Pak in the same direction and rotating the sample stage after placing the bottom BSCCO flake. In the later stage of sample fabrication, two BSCCO flakes on the same Gel-Film was placed on the substrate sequentially, which guarantees better angular alignment. The samples were annealed for typically 10~minutes in either flowing oxygen (ambient pressure) at 530$^\circ$C or in ozone (around $5\times10^{-5}$~mbar) at 450$^\circ$C. Ozone gas was introduced into the annealing chamber from a home-built liquid ozone system (99\% purity). As a further improvement, we later attached the ozone annealing chamber directly to the glovebox. The junctions can be annealed in this upgraded system without exposing to air, giving rise to a much higher yield. The annealed samples were all wired to the chip carriers within 0.5-1 hour at ambient condition and loaded into the transport measurement system (Oxford TeslatronPT, 1.5-300~K). For resistance measurements, we employed either the standard lock-in technique (Stanford Research SR830) with an ac current of 1~$\mu$A at 30.9~Hz or a dc measurement technique offered by the combination of the current source and nanovoltmeter--Keithley~6221/2182A (Delta mode). For tunneling measurement, we used Keithley~6221 for sending the current and Keithley~2182A for measuring the voltage in a four-terminal configuration. The TEM samples were prepared by using focused ion beam (FIB) technique. TEM experiments were performed on an aberration-corrected ARM200CF (JEOL), operated at 200~keV and with double spherical aberration (Cs) correctors, which usually renders a spatial resolution of 0.8~$\overset{\circ}{\rm A}$.

\section{JUNCTION GEOMETRY and ATOMIC STRUCTURE}

Stacking ultrathin BSCCO flakes produces ideal $c$-axis Josephson junctions with macroscopically uniform and atomically sharp interfaces. We demonstrate this by a series of TEM images in Fig.~2. The TEM images were taken from three samples with different twist angles. Based on the intensity contrast, we can identify the exact location of the artificial interface and thus color-code the top and bottom BSCCO flakes, as well as the SiO$_2$/Si substrate. The lateral size of the image in Fig.~2(a) is 8~$\mu$m, thus it reveals a macroscopically uniform interface that extends over the whole junction. This uniformity assures that our tunneling experiment is close to the situation conceived by theory, as will be discussed in the next sections. Consecutive zoom-in images of the same sample are shown in Fig.~2(b) and Fig.~2(c), where the interface between the top and bottom BSCCO flakes (Fig.~2(b)) and the individual BiO layers imaged as bright stripes (Fig.~2(c)) can be clearly resolved. The perfect junction structure we obtained are also illustrated by the TEM images in Fig.~2(d) and Fig.~2(e) taken on other two samples, and the uniform nature of the junctions goes down to the atomic level (the Bi atoms in the BiO layer are imaged as individual bright spots in Fig.~2(e)). In short, our junctions are of superb homogeneity over the entire tunneling junction region. We stress that such quality of the junctions is unprecedented and has not been reported before.

Since the samples with 45$^\circ$ twist angle are crucial for distinguishing $s$ and $d$-wave pairing---the main aim of this work, we characterize their interface structure in more details. Fig.~3 exhibits two atomically resolved images taken from the same area of a junction with a nominal twist angle of 45$^\circ$, and we can see the individual Bi atoms of the BiO layers (indicated by the black arrows). The top and bottom flakes are viewed from the [1$\overline{1}$0] and [010] directions, respectively. The twist angle deviates slightly from 45$^\circ$ such that only half of the junction can be atomically resolved each time. The well-known super-modulation can also be clearly resolved for the top BSCCO flake and it persists all the way down to the interface. We observe in Fig.~3 three BiO layers (indicated by the red arrows) in the twist junction: two supermodulated layers from the top flake and one from the bottom flake. It seems that a sizable fraction of the barrier at the interface consists of triple BiO layers, which is different from the usual case of a BiO bilayer in bulk BSCCO (more data are shown in Fig.~S1 of the supplementary information).

To determine the error in the twist angle experimentally, we totally prepared three $\theta_\mathrm{nominal}=45^\circ$ junctions and investigated them by TEM. We characterize this error by measuring the rotation angle of the sample stage that was necessary to rotate from the position for clearly resolving BSCCO-1 to that for BSCCO-2. If there is no error in the twist angle, BSCCO-1 and BSCCO-2 should be simultaneously resolved by TEM. We obtained 2.2$^\circ$, 2.7$^\circ$ and 3.1$^\circ$ for three samples, respectively. We double checked these angles with another method. We obtain two Kikuchi patterns by aligning the electron beam with BSCCO-1 and BSCCO-2, respectively. The distance between the centers of the two Kikuchi patterns can be then converted to the deviation angle, since we know the convergence angle of the transmission electron beam spot. This method yields 2.0$^\circ$, 3.0$^\circ$ and 3.3$^\circ$ for the three samples, respectively. Based on the consistent measurements, the error in the twist angles is determined to be about $\pm3^\circ$.

\begin{figure}
\includegraphics[width=86mm]{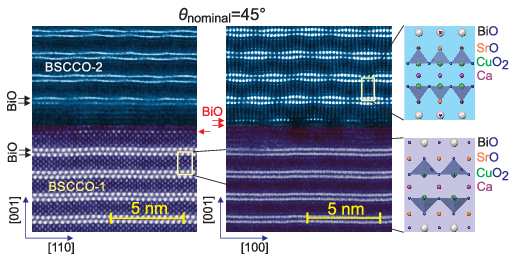}
\caption{Atomically resolved TEM images around the BSCCO-1/BSCCO-2 interface. Arrows mark the BiO layers. Right panels illustrate the atomic structure of 0.5 UC BSCCO viewed from [010] (upper) and [1$\overline{1}$0] (lower) directions, respectively.
  }
\end{figure}

\section{Josephson tunneling}

\begin{figure}
\includegraphics[width=86mm]{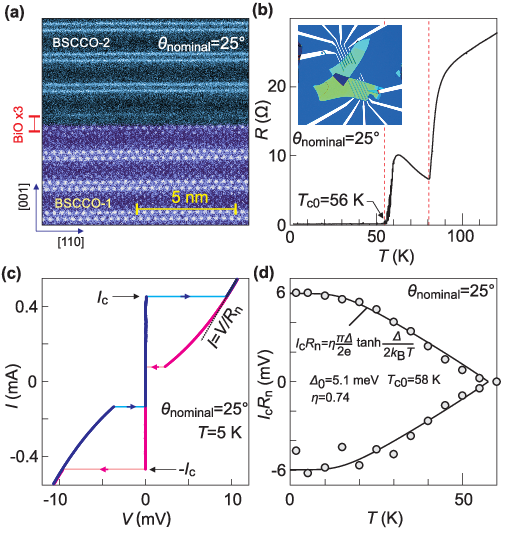}
\caption{
  Data of a BSCCO/BSCCO junction with a nominal twist angle of 25$^\circ$. (a) High resolution transmission electron microscope image of the junction region taken after the low temperature transport measurements. (b) Temperature dependent resistance across the junction. The first drop around 80 K stems from the superconducting transition of the individual BSCCO flakes. The second drop around 56 K is the superconducting transition in the artificial junction. Inset shows an optical image of the sample. (c) Tunneling $I-V$ characteristic measured across the artificial junction. Blue and pink colors represent the positive and negative sweep directions, respectively. (d) $I_cR_n$ as a function of temperature for both the positive and negative biases. Solid curves are fits to a modified Ambegaokar-Baratoff formula (shown in the inset). The fitting parameters are listed in the figure.
  }
\end{figure}

With the atomic structure of our junctions clarified, we demonstrate the Josephson tunneling by using the sample with a nominal twist angle $\theta_\mathrm{nominal}=25^\circ$ first. Fig.~4(a) shows the high resolution TEM image of the junction, which was obtained after the transport measurements. It confirms that the atomic structure at the artificial interface remains intact. The temperature dependent resistance across the junction is shown in Fig.~4(b). We observe two distinct transitions: the first transition occurs at around 80~K, at which the individual flakes become superconducting; the second transition occurs at 56~K, indicating that the junction region becomes superconducting at this temperature. We ascribe the latter to a lower doping level at the junction region and thus a lower superconducting transition temperature than that of the rest of the flakes. We further confirm this point by separately measuring the interlayer and intralayer transport on other samples. The lower doping at the interface may be caused by two factors: (1) oxygen loss at the cleaved surface during the exfoliation; (2) extra charge transfer due to the triple BiO layer at the interface.

Shown in Fig.~4(c) is the $I$-$V$ characteristic of this sample measured at $T$ = 5~K, i.e., $T$~\textless~0.1$T_c$. The reduced size of our samples, in comparison to the bulk bicrystal~\cite{Li_1999}, allows us to perform tunneling experiments down to low temperatures without current heating effect. We have carefully checked that the hysteresis seen here is not caused by Joule heating since we obtain the same $I$-$V$ characteristic with different current sweeping rates or sweeping ranges. The Josephson current flows without dissipation across the junction, giving rise to the vertical line in Fig.~4(c). Notably, sweeping the voltage in one direction leads to only two switching points between the dissipationless state and the resistive one (horizontal arrows). This is in sharp contrast to the multiple branches seen in the vertical transport of BSCCO Josephson junctions~\cite{Takano_2002,Latyshev_2004,Suzuki_2000,Suzuki_2012}, where the number of branches reflects the number of tunneling junctions involved. The missing of multiple branches here stems from the fact that the twist junction becomes resistive at a critical current well below that within the BSCCO flakes. Transport across the two flakes is therefore dominated by the tunneling between the top 0.5~UC in BSCCO-1 and the bottom 0.5~UC in BSCCO-2. This is one of the unique features of the Josephson junctions studied in our experiment.

In Fig. 4(d), we illustrate the temperature dependence of $I_c R_n$. $I_c$ is the Josephson critical current [marked in Fig.~4(c)] and $R_n$ is the normal state resistance obtained from the slope of the $I$-$V$ curve at high bias voltage. As seen in Fig.~4(d), $I_c R_n$ drops to zero at a temperature at which the individual flakes are still superconducting [Fig.~4(b)], further supporting the aforementioned assignment of the Josephson coupling to the artificial interface. The solid curves in Fig.~4(d) are fits to a modified Ambegaokar-Baratoff formula~\cite{Ambegaokar_1963,Ambegaokar_1963b,Xu_1994}~[listed inside Fig.~4(d)]:
 \begin{equation}
   I_c R_n=\eta \frac{\pi\Delta(T)}{2e}\tanh{\frac{\Delta(T)}{2k_BT}},
 \end{equation}
 where $\Delta(T)=\Delta_0\tanh(1.74\sqrt{T_{c0}/T-1})$ and $\Delta_0$, $T_{c0}$ and $\eta$ are fitting parameters. This simple formula seems to satisfactorily capture the experimental features. The fitted $\Delta_0$$=$5.1~meV is smaller than the value of $1.764k_BT_{c0}=8.8$~meV---from the standard BCS ratio, which may reflect the asymmetric superconduting gaps on the two sides of the junction~\cite{Xu_1994}. The prefactor $\eta=0.74$ represents the quantitative deviation from the original Ambegaokar-Baratoff formula. Its reduction from unity may be attributed to either the presence of $d$-wave pairing~\cite{Xu_1994} or varying degree of tunneling coherence between $s$-wave superconductors~\cite{Klemm_1998}. The temperature dependence therefore may not yield decisive information on the pairing symmetry. Nevertheless, these results demonstrate that our junction behaves largely as expected from the Ambegaokar-Baratoff theory and the transport quality is comparable to those intrinsic junctions~\cite{Suzuki_2012}.

To further demonstrate that our junctions are Josephson coupled, we carry out the in-plane magnetic field ($B_\parallel$) study of $I_c$. Realizing the standard Fraunhofer pattern requires a small junction size $W$, in comparison to both the $c$-axis London penetration depth $\lambda_c$ and the Josephson penetration depth $\lambda_j$~\cite{Kleiner_1994}. $\lambda_c$ ranges from tens of micrometers to a few hundred of micrometers while $\lambda_j$ is as small as a few hundred nanometers~\cite{Kleiner_1994}. Fig.~S2 of the supplementary information shows that $I_c(B_\parallel)$ in our relatively large junctions shows the non-monotonic behavior but departs from the ideal Fraunhofer pattern. In order to approach the small-junction limit, we employ focused ion beam to cut out a small area from a large junction. Fig.~5 shows the data from these junctions. We defer further discussions on the observation of Josephson coupling at 45$^\circ$ to the following paragraph. Here, we point out that Fig.~5 clearly indicates the Fraunhofer pattern of $I_c(B_\parallel)$. Notably, $I_c(B_\parallel)$ in Fig.~5(a) displays complete suppression of $I_c$ at the first, second and third nodes (indicated by arrows). The solid curves in Fig.~5 are based on the simple formula for the Fraunhofer pattern: $I_c=I_{c,0}|\sin\Phi/\Phi|$, where $\Phi=\pi B_\parallel/B_0$. $I_{c,0}$ and $B_0$ are fitting parameters. Further analysis of $B_0$ is given in section III of the supplementary information.

\begin{figure}
\includegraphics[width=86mm]{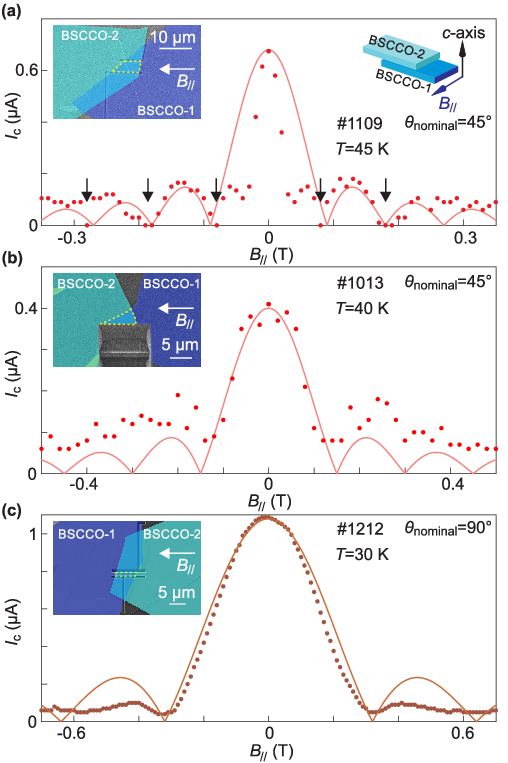}
\caption{
  Josephson critical current as a function of the magnetic fields for three junctions with $\theta_\mathrm{nominal}=45^\circ$ (a,b) and $90^\circ$ (c), respectively.
  $B_\varparallel$ indicates the field applied parallel to the BSCCO flakes. Solid curves are the Fraunhofer pattern. Insets show the false colored scanning electron microscope images. Dotted lines demarcate the junction regions carved out by focused ion beam.
  }
\end{figure}

\begin{figure}
\includegraphics[width=86mm]{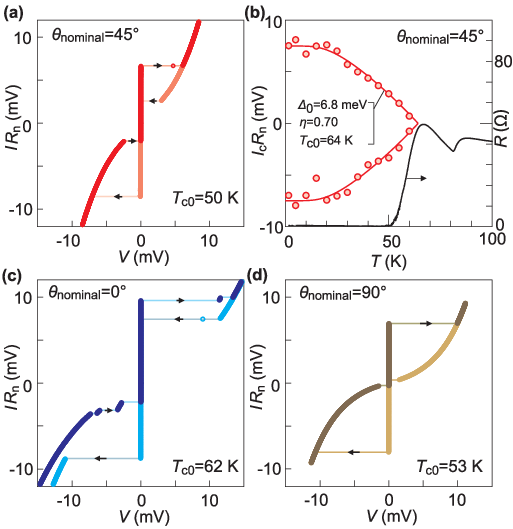}
\caption{
  (a)(c)(d) Normalized tunneling characteristics of BSCCO junctions at different twist angles (measured at 1.6~K). $T_{c0}$ represents the transition temperature of the junction region, as obtained from the temperature-dependent resistance measurement. (b) $I_cR_n$ as a function of temperature for both the positive and negative biases for the sample with $\theta_\mathrm{nominal}=45^\circ$. Red curves are fits to a modified Ambegaokar-Baratoff formula. Black curve is the temperature dependent resistance measured across the junction. The temperature where $I_c R_n$ drops to zero is slightly higher than $T_{c0}$ from resistance measurement, indicative of disconnected superconducting puddles at $T>T_{c0}$.
  }
\end{figure}

Now, we move to discuss the results obtained from the $\theta_\mathrm{nominal}=45^\circ$ twist junctions. In order to compare the data from different samples, we normalize the current axis by multiplying $R_n$ [Fig.~6(a)]. The temperature dependence of $I_c R_n$ can be again fitted by Eq.~(1) [Fig.~6(b)]. To our surprise, as shown in Fig.~6(a) and Fig.~6(b), the sample exhibits $I_c R_n$ value of 7.7~mV, which is even a bit larger than that (5.5~mV) at $\theta_\mathrm{nominal}=25^\circ$ [Fig.~5(d)]. The presence of the sizable Josephson tunneling at $\theta_\mathrm{nominal}=45^\circ$ directly contradicts the expectation from the established theories that consider purely $d$-wave pairing symmetry~\cite{Klemm_1998,Arnold_2000,Bille_2001,Klemm_2003,Klemm_2005,Yokoyama_2007}. Since it is the most important result of this study, here we give more discussion about the theoretical argument.

The angular dependent Josephson current between two rotated CuO$_2$ planes---especially its full suppression at a twist angle of 45$^\circ$---is a manifestation of the sign-change property of the $d$-wave pairing symmetry~\cite{Klemm_1998,Arnold_2000,Bille_2001,Klemm_2003,Klemm_2005,Yokoyama_2007}. We recapitulate the theory~\cite{Klemm_2005} using the schematic in Fig.~7. Details of the calculations used to analyze our data are given in the supplementary information. Fig.~7(c) represents the tunneling component when the top CuO$_2$ plane is rotated clockwise by 45$^\circ$: $\langle F_{t-b} G_t(45^\circ)G_b\rangle$,
where we integrate over each of the two first Brillouin zones, $F_{t-b}$ represents the squared modulus of the tunneling matrix element, $G_t$ and $G_b$ are the anomalous Green functions which are proportional to the order parameters of the top and bottom layers, respectively. In contrast, rotating the top CuO$_2$ in the opposite direction by 45$^\circ$ [Fig.~7(b)] gives rise to a tunneling component of $\langle F_{t-b} G_t(-45^\circ)G_b \rangle$ such that
\begin{equation}
  \langle F_{t-b}G_t(45^\circ)G_b\rangle = -\langle F_{t-b}G_t(-45^\circ)G_b\rangle.
\end{equation}
The minus sign in front of the term on the right side of Eq.~(2) reflects that the order parameter changes sign across the nodes. For $s$-wave order parameter, there is no minus sign, i.e. rotating clockwise and anticlockwise by 45$^\circ$ are equal. Viewing the latter situation [Fig.~7(b)] from the backside gives rise to the situation shown in Fig.~7(d), which is obtained by simply exchanging the indices:
\begin{equation}
  \langle F_{t-b}G_t(-45^\circ)G_b\rangle = \langle F_{b-t}G_b(-45^\circ)G_t\rangle.
\end{equation}
By further rotating the whole system clockwise with an angle of 45$^\circ$, Fig.~7(d) can be immediately transformed to Fig.~7(c) as follows:
\begin{equation}
   \langle F_{b-t}G_b(-45^\circ)G_t\rangle=\langle F_{b-t} G_b G_t(45^\circ)\rangle.
\end{equation}
The right side of Eq.~(4) is identical to $\langle F_{t-b}G_t(45^\circ)G_b\rangle$ because: $\langle F_{t-b} \cdot\cdot\cdot \rangle$ can be switched to $\langle F_{b-t} \cdot\cdot\cdot \rangle$, since we integrate over the whole Brillouin zone of each layer~\cite{Note01}. Combining Eq.~(2) to Eq.~(4), we obtain $\langle F_{t-b}G_t(45^\circ)G_b\rangle = -\langle F_{t-b}G_t(45^\circ)G_b\rangle$, thus the tunneling component has to be zero. Apparently, the observation of a finite $I_c R_n$ value at $\theta_\mathrm{nominal}=45^\circ$ runs directly against the analysis above for a pure $d$-wave pairing symmetry.

\begin{figure}
\includegraphics[width=86mm]{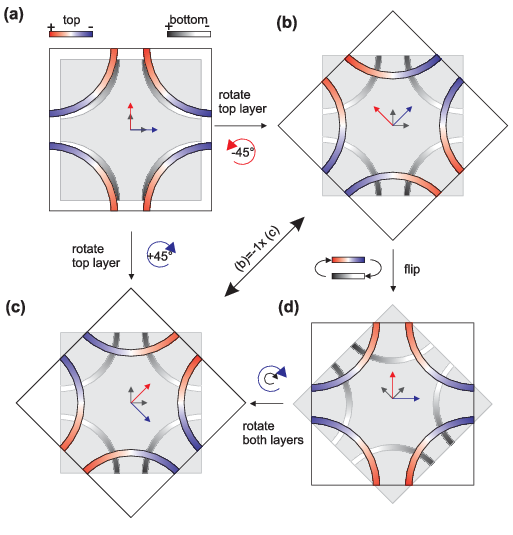}
\caption{
  (a) Illustration of the CuO$_2$ bands in the first Brillouin zone around the $\Gamma$ point. The top (bottom) layer of CuO$_2$ is represented by blue and red (black and white) colors. The change from blue (black) to red (white) reflects the sign change of the order parameter.
  (b)(c) Tunneling element in the two cases where the top CuO$_2$ layer is rotated clockwise or counterclockwise by 45$^\circ$. The minus sign between them indicates that the situation of (b) transforms to that of (c) by multiplying a factor of -1.
  (d) This tunneling situation is obtained by viewing the tunneling element shown in (c) from the backside. It transforms back to (b) by rotating the top and bottom layer together by 45$^\circ$.
  }
\end{figure}

In reality, one may argue that the twist angle can depart from 45$^\circ$ by $\sim\pm3^\circ$ as determined by TEM, and this deviation may allow a finite supercurrent even for $d$-wave pairing symmetry. In order to rule out this possibility, we compare the data obtained at $\theta_\mathrm{nominal}=45^\circ$ with those obtained at $\theta_\mathrm{nominal}=0^\circ$ and 90$^\circ$ [Fig.~6(c), Fig.~6(d)]. Clearly, they show similar $I_c R_n$ values. For a more quantitative comparison, we use the expected $\cos(2\theta)$ dependence for the tunneling between $d$-wave superconductors. By using the $I_c R_n$ values (9.2~mV$/$7.5~mV) at 0$^\circ$$/$90$^\circ$, we obtain that a junction with $\theta$ = 48$^\circ$ or 42$^\circ$ can possess $I_c R_n$ of $\sim$0.8 to 1~mV. Apparently, this upper bound is well below the measured value at $\theta_\mathrm{nominal}=$45$^\circ$ [Fig.~6(a)]. The angular deviation can be therefore excluded as the mechanism for the persistent Josephson coupling at the twist angle of 45$^\circ$. We will discuss the implication of this result in the final section.

\section{Data analysis}
\begin{figure}
\includegraphics[width=86mm]{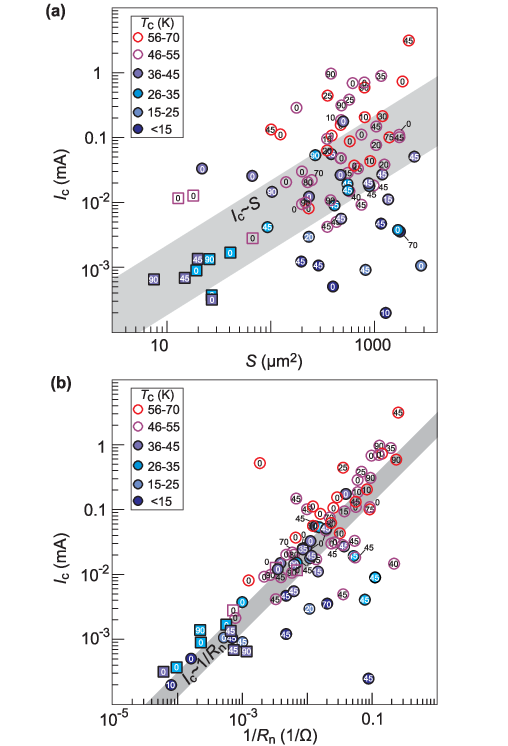}
\caption{(a) Josephson critical current $I_c$ as a function of the physical area of the junction. (b) $I_c$ as a function of the inverse normal state resistance $1/R_n$. All data points are obtained from $I-V$ measurements at 1.6 K. Numbers inside or close to the symbols indicate the nominal twist angles. Thick gray band-like backgrounds in (a)(b) represent the linear dependence. Squares represent data obtained from samples with small junction areas, which are realized by focused ion beam etching.
  }
\end{figure}

\begin{figure}
\includegraphics[width=86mm]{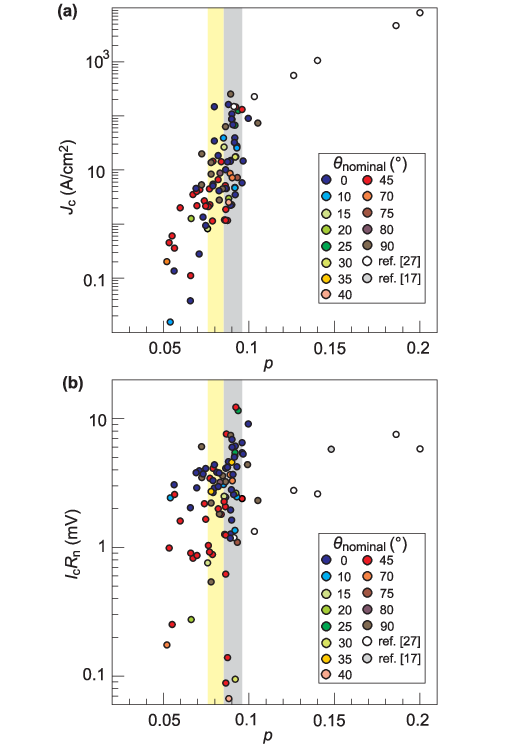}
\caption{
$J_c$ (a) and $I_c R_n$ (b) as a function of doping for junctions with different twist angles. The doping $p$ is estimated from the corresponding $T_{c0}$ by using the empirical formula: $T_{c0}=91\cdot[1-82.6(p-0.16)^2]$. The empty circles correspond to tunneling in intrinsic Josephson junctions of a single BSCCO, as obtained from ref.~\cite{Suzuki_2012}. The gray circle is from previous 45$^\circ$ twist bicrystal experiments by using bulk crystals of BSCCO at optimal doping\cite{Li_1999}.
  }
\end{figure}

\begin{figure*}
\includegraphics[width=178mm]{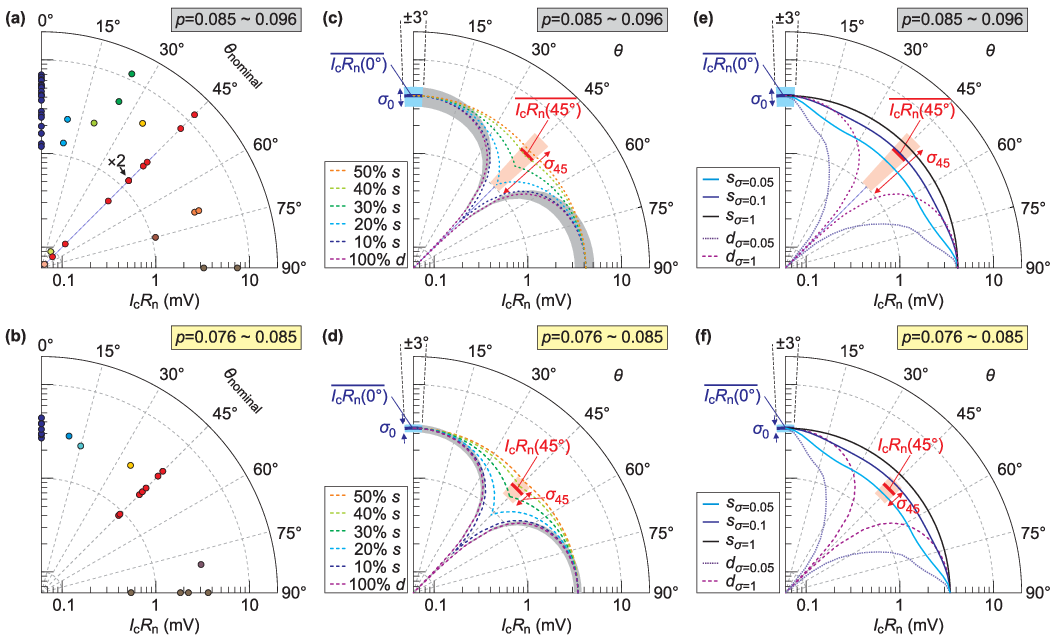}
\caption{
(a)(b) Fan chart diagram of $I_c R_n$ values as a function of the nominal twist angle for BSCCO junctions. The range of doping levels are indicated as shaded regions in Fig.~9. (c)(d) Comparison between $\overline{I_c R_n}$ values and the theoretically calculated angular dependence with $s$-wave/$d$-wave mixing. The error of $\pm$3$^\circ$ in twist angles is estimated from the TEM experiments. The dashed curves represent the theoretically expected angular dependence from purely $d$-wave to 50\% mixture of $s$-wave (50\% $s$). The shaded gray region indicates the purely $d$-wave dependence by taking into account the data spread of $\sigma_0$. (e)(f) Comparison between the experimentally obtained $\overline{I_c R_n}$ and the theoretically calculated angular dependence with different degrees of tunneling coherence (varied by the parameter $\sigma$, as explained in the supplementary information). The theoretical curves are normalized by multiplying the calculated critical current $J_c(\theta)$ with a factor: $\overline{I_c R_n(0^\circ)}/J_c(0^\circ)$.
  }
\end{figure*}

To further understand the Josephson coupling in $c$-axis twist junctions, we carry out an extensive study by collecting data from over 90 samples. Figure~8(a) provides an overview of $I_c$ as a function of the junction size $S$. They in general follow the $I_c\propto S$ scaling. Figure~9(a) provides further insight into the doping dependence of $J_c$, calculated from $I_c/S$. In general, $J_c$ becomes larger with increased doping. This increasing trend connects smoothly to the doping behavior of the Josephson tunneling in single crystalline BSCCO~\cite{Suzuki_2012} (empty circles). This agreement demonstrates that the quality of our stacked junction is as high as the intrinsic ones.

One factor that contributes to the data spread in Fig.~8(a) is the deviation of the actual tunneling area from $S$, because the region close to the edge of the flake may be non-superconductive due to prominent oxygen loss~\cite{Truccato_2005}. We therefore employ $I_c R_n$ to represent the Josephson coupling strength. Both $I_c$ and $1/R_n$ are proportional to the actual tunneling area, as supported by the linear dependence between them [Fig.~8(b)], such that this sample dependent parameter cancels out in $I_c R_n$. Fig.~9(b) plots the doping dependence of $I_c R_n$. Notably, our data is also consistent with that obtained in the previous experiment by using bulk bicrystal~\cite{Li_1999}, demonstrating again the high quality of the junction.

We focus on the data in two narrow doping ranges [shaded gray/yellow in Fig.~9(b)] and plot them as a function of the twist angle in Fig.~10(a)(b) as fan chart diagrams. Here the vertical axis represents a twist angle of zero degree. By rotating clockwise around the origin, the twist angle changes from 0 to 90 degrees (horizontal axis). The radius of each data point from the origin indicates the magnitude of the $I_c R_n$ value. Fig.~10(a)(b) show that the sample-to-sample variation is predominantly on the same order of magnitude. The corresponding $T_c$, as determined from temperature dependent resistances, varies from 49.5~K to 60.5~K for samples shown in Fig.~10(a) and from 38.5~K to 49.5~K for those in Fig.~10(b), respectively. The possible variation in barrier thickness from a double BiO layer to a triple BiO layer (Fig.~S1) cannot account for the data spread because different samples show similar distribution of the number of BiO layers. Some other factors must be playing an important role in causing the spread of $I_c R_n$.

We discuss the situation at $\theta_{nominal}=0^\circ$ first [blue circles in Fig.~10(a)(b)] since any intrinsic angular dependence, for example the $\cos (2\theta)$ behavior expected for $d$-wave junctions, should be negligible within the experimental precision ($\pm3^\circ$) around 0$^\circ$. As our samples are in the mesoscopic regime and were annealed individually, the doping level at the interface may vary at different locations, which fluctuates from sample to sample. For two samples showing the same maximum $T_c$, which corresponds to $T_c$ observed in transport, the sample with a broader distribution of doping would generate a smaller $I_c R_n$ because the contribution from the high doping portion is smaller. Another effect that accompanies the doping inhomogeneity is the varying degree of tunneling coherence. The interstitial oxygen in BSCCO, which is crucial for the carrier doping, also serves as scattering centers. The doping fluctuation can therefore lead to different degree of momentum preservation in the tunneling process. These factors cause the Josephson coupling strength to be sample dependent. In Fig.~10(c)-(f), we quantify the sample to sample variation by averaging over the data points at $0^\circ$. We plot $\overline{I_c R_n(0^\circ)}$ and the sampling standard deviation ($\sigma_0$) as the dark blue bars and light blue shades, respectively.

Despite the uncertainty brought by the doping inhomogeneity, the symmetry argument for a $d$-wave superconductor as sketched in Fig.~7 can still be applied locally. Even if Josephson tunneling occurs in multiple parallel regions with different doping levels, one can sum them up as $\sum_m \langle F_{t-b}^m G_t^m(-45^\circ)G_b^m \rangle$, where we index the different regions by an integer--$m$. The doping levels determine the absolute values of $G_t^m$ and $G_b^m$ in Eqs.~(2)-(4) while the varying degree of tunneling coherence is captured by the different magnitude of $F_{t-b}^m$. They do not affect the sign-change property imposed by $d$-wave symmetry and the Josephson coupling should still drop to zero at 45$^\circ$. Based on the fact that the Josephson coupling strength at 0$^\circ$ spreads around $\overline{I_c R_n(0^\circ)}$ by $\sigma_0$ due to the doping uncertainty, we expect $I_c R_n$ at other twist angles to exhibit a data spread that crowd within the shaded gray regions in Fig.~10(c)(d), because the upper and lower bounds set at $0^\circ$ should also follow the $\cos(2\theta)$ dependence. Clearly, $\overline{I_c R_n (45^\circ)}$ together with $\sigma_{45}$ are outside the demarcated region.

We provide an explanation for the large $\sigma_{45}$ shown in Fig.~10(c). This large data scatter is strongly influenced by the data points in Fig.~10(a) with $I_cR_n$ around 0.1~mV. In Section IV of the supplementary information, we give a historical account of our data. Those data points in Fig.~10(a) around 0.1~mV in fact originate from samples fabricated either in the early stage or in the beginning when a new recipe was used. Some extrinsic contamination might be present and caused the lower Josephson coupling. Excluding this additional technical factor would raise the averaged value and substantially reduce the data scatter of $\sigma_{45}$ in Fig.~10(c).

From Fig.~10(c)(d), we conclude that, rather than the doping inhomogeneity and disorder induced scattering, the disagreement between our data and the theoretical expectation for a purely $d$-wave superconductor has an intrinsic origin. With both the atomic structure and the Josephson tunneling carefully elucidated, the experimental results indicate the possible presence of an $s$-wave order parameter.

Based on the phenomenological model in ref.~\cite{Klemm_2005}, we consider two scenarios--both involving an anisotropic $s$-wave order parameter--to account for the persistent Josephson coupling at the twist angle of 45$^\circ$. The calculation details are given in Section I of the supplementary information. Fig.~10(c)(d) show the theoretically calculated angular dependencies when the two neighboring superconductors possess an order parameter mixed of $d$-wave and $s$-wave. By comparing the theoretical curves with the experimental data, the $s$-wave component seems to take a noticeable portion from 20-30\%, based on the lower bound of $\overline{I_c R_n(45^\circ)}-\sigma_{45}/2$, to as high as 40\%, if only the averaged values are compared.

In the second scenario, the slight reduction of $\overline{I_c R_n(45^\circ)}$ from $\overline{I_c R_n(0^\circ)}$ can be accounted for by the orbital effect. In the momentum conserved situation i.e. coherent tunneling, the rotation changes the overlapping portion of the two Fermi surfaces. Even the Josephson tunneling between two $s$-wave superconductors can exhibit angular dependence~\cite{Bille_2001}. In Fig~10(e)(f), we show the theoretical situations with different degrees of tunneling coherence. By increasing the degree of incoherence (theoretically parameterized as $\sigma$), this angular dependence quickly diminishes, giving rise to a fully isotropic behavior for $s$-wave superconductors [solid black curve in Fig.~10(e)(f)]. Our experimental data seems to fall in the range where $\sigma$ varies by one order of magnitude for a purely $s$-wave junction. Notably, increasing the tunneling coherence between two $d$-wave superconductors [dotted curve in Fig.~10(e)(f)] would only result in a larger disagreement with experiment.

\section{Discussion}

We find that the Josephson coupling strength between two BSCCO blocks seems to host a significant portion of $s$-wave order parameter, which challenges the present theoretical understanding based on pure $d$-wave pairing symmetry~\cite{Klemm_1998,Arnold_2000,Bille_2001,Klemm_2003,Klemm_2005,Yokoyama_2007}. In the following, we discuss the influence of structural distortion and interfacial symmetry breaking, which may give rise to an emergent $s$-wave pairing.

First, the exfoliated thin flakes may suffer from surface roughness of the substrate~\cite{Dean_2010}.  However, a recent STM study~\cite{Yu_2019} on a monolayer of BSCCO flake (0.5~UC) exfoliated on the SiO$_2$/Si substrate reveals the same tunneling spectrum as that obtained from bulk samples. The substrate seems to play a minor role in determining the electronic property of ultrathin BSCCO flakes. Notably, our BSCCO is about 10-20 UC thick such that the distortion brought by possible bumps on the substrate should be further smeared.

Secondly, there may exist variation of the tunnel barrier thickness, as seen in TEM images (Fig.~S1). It is conceivable that local distortions exist in regions where the tunnel barrier transits from double BiO layers to triple BiO layers. These regions may allow the emergence of $s$-wave pairing symmetry. However, the distorted region of a few nanometers takes only a small fraction of the junction extending several micrometers (Fig.~S1), which is insufficient to account for the significant portion of $s$-wave as indicated in the previous section.

Thirdly, the twisted junction may spontaneously break the four-fold rotational symmetry and give rise to an in-plane anisotropy of the electronic states. In YBa$_2$Cu$_3$O$_7$ (YBCO), an $s$-wave component caused by the anisotropic electronic state has been identified, which is reported to take 15\% portion of the order parameter~\cite{Smilde_2005}. Although the 45$^\circ$-twist BSCCO/BSCCO junction satisfies the C4 symmetry, in reality a junction may deviate slightly from the designated angle. A junction with a $44^\circ$ twist angle, for example, can effectively break the C4 symmetry, which may result in a mixture of $d$-wave and $s$-wave order parameters. However, our experimental observation of a persistent $s$-wave order parameter would indicate that the twisted BSCCO (BSCCO-2, for example) influences on the neighboring BSCCO (BSCCO-1) in a similar fashion as the orthorhombic distortion in YBCO. This analogy seems counter-intuitive as the coupling between each 0.5~UC of BSCCO is the weak van der Waals interaction.

Fourth, so far theories dealing with the $c$-axis tunneling considers a symmetric situation with a simple tunneling matrix element~\cite{Klemm_1998,Arnold_2000,Bille_2001,Klemm_2003,Klemm_2005,Yokoyama_2007}.
Due to the presence of impurities and the additional BiO layer in the barrier, the tunneling matrix element may be more complicated. From symmetry point of view, the triple BiO barrier, as we found in our junctions, allows the introduction of an antisymmetric tunneling matrix element, in addition to the symmetric one. In this case, the squared modulus of the tunneling matrix element may not be symmetric, such that  $\langle F_{t-b} \cdot\cdot\cdot \rangle$ discussed in Eqs.~(2)-(4) may not be identical to $\langle F_{b-t} \cdot\cdot\cdot \rangle$. However, we note that the asymmetric term usually enters the Hamiltonian as a second-order term via the spin-orbit interaction. Given the fairly weak spin-orbit interaction in Cu and O derived bands, it is unlikely that such a symmetry breaking term plays a major role.

To summarize, none of the scenarios above satisfactorily explains the large Josephson tunneling observed at $\theta_{nominal}=45^\circ$ at the moment, calling for further theoretical investigations. Given the well-defined junction structure with atomic-level control here, this non-trivial feature can impose further constraints on a microscopic model for the high temperature superconductivity in BSCCO. The van der Waals stacking technique demonstrated here can also be extended to investigate the $c$-axis tunneling in other high temperature superconductor~\cite{Zalic_2019}.

Recently, a time reversal symmetry breaking $d\pm id^\prime$ or $d\pm is$ order parameter was proposed to emerge in the 45$^\circ$-twist double CuO$_2$ layers~\cite{Can_2021}. The formation of such high temperature topological superconductivity may account for the large Josephson coupling. It may also lead to a U-shaped gap in the 45$^\circ$ twist junction, which is in direct contrast to the widely seen V-shaped gap in untwisted junctions~\cite{Krasnov_2000,Suzuki_2000}. While more experiments are necessary to draw a definite conclusion, our present data may provide some clue to the relevance of this interesting proposal. Especially, the finite slopes of the left and right branches in the $I$-$V$ characteristics essentially reflect the density of states inside the superconducting gap~\cite{Krasnov_2000,Suzuki_2000}. These sloped branches seem to show qualitatively the same behavior in junctions with different twist angles (see Fig.~6 and extended data in Fig.~S6-S8). Furthermore, the temperature dependence of these branches seem to be qualitatively the same at 0$^\circ$ and 45$^\circ$ twist angles (Fig.~S5). These experimental features seem to be different from the theoretical expectation. However, we so far focus on the small energy range for the Josephson effect and our junctions still consist of a few UC of BSCCO on each side, which is not in the 0.5 UC-0.5 UC limit considered in the theory. Further endeavors are underway to carry out the tunneling spectroscopy in a wider energy scale and to thin down the junction to the monolayer limit.

\begin{acknowledgments}
We thank Yijun Yu and Yuanbo Zhang for generous help in sample fabrications, and Yang Li and Yong Xu for technical assistance in theoretical calculations. We acknowledge insightful discussions with Hong Yao and Haiwen Liu. This work is financially supported by the Ministry of Science and Technology of the People's Republic of China (2017YFA0302902, 2017YFA0304600,2018YFA0305603); the National Natural Science Foundation of China (grant No. 11790311, 51788104, 11922409, 12074039, 12004041) and the Beijing Advanced Innovation Center for Future Chips (ICFC). Work at Brookhaven is supported by the Office of Basic Energy Sciences, Division of Materials Sciences and Engineering, U.S. Department of Energy under Contract No. DE-SC0012704.
\end{acknowledgments}

\bibliographystyle{apsrev4-1}

\end{document}